\documentclass{article}
\usepackage{amsmath}
\usepackage{graphicx}
\usepackage[utf8]{inputenc}
\usepackage{hyperref}
\usepackage{geometry}
\usepackage[numbers]{natbib}
\title{Establishing Workload Identity for Zero Trust CI/CD: From Secrets to SPIFFE-Based Authentication}
\author{
Surya Teja Avirneni \\
Cloud Platform and Security Architect, \\
IEEE Member, ISC2 Certified, ACM Member \\
United States
}
\date{April 2025}

\begin{document}

\maketitle

\begin{abstract}
CI/CD systems have become privileged automation agents in modern infrastructure, but their identity is still based on secrets or temporary credentials passed between systems. In enterprise environments, these platforms are centralized and shared across teams, often with broad cloud permissions and limited isolation. These conditions introduce risk—especially in the era of supply chain attacks—where implicit trust and static credentials leave systems exposed.

This paper describes the shift from static credentials to OpenID Connect (OIDC) federation, and introduces SPIFFE (Secure Production Identity Framework for Everyone) as a runtime-issued, platform-neutral identity model for non-human actors. SPIFFE decouples identity from infrastructure, enabling strong, portable authentication across job runners and deployed workloads. We show how SPIFFE identities support policy alignment, workload attestation, and mutual authentication. The paper concludes by outlining next steps in enabling policy-based access, forming the basis of a broader Zero Trust architecture for CI/CD.
\end{abstract}

\section{Introduction}

Continuous Integration and Continuous Deployment (CI/CD) platforms are essential automation layers in modern software delivery. These systems orchestrate privileged workflows—building, testing, and deploying code—but are often treated as anonymous infrastructure actors. Historically, CI/CD pipelines have relied on static secrets, shared service accounts, and injected credentials to interact with downstream systems.

This approach creates significant risk. In many enterprises, CI systems are multi-tenant, centrally managed platforms with broad access to cloud resources. A compromised CI job can lead to privilege escalation, cross-tenant data access, or unauthorized production changes. The rise of software supply chain attacks—such as dependency hijacking or malicious release tampering—further exposes the weakness of identity assumptions in pipeline automation.

Zero Trust architecture requires strong identity and continuous verification for all entities, including non-human actors. However, common patterns such as OIDC federation remain tightly bound to CI platform metadata and are difficult to federate or audit across environments. SPIFFE (Secure Production Identity Framework for Everyone) addresses this by issuing cryptographically verifiable identities to workloads at runtime, based on attested selectors such as container metadata, service accounts, or host attributes.

This paper focuses on the application of SPIFFE to CI/CD workflows. We examine the evolution from secrets and OIDC federation to runtime-issued identity, describe how SPIFFE integrates with ephemeral jobs and target workloads, and demonstrate how this model enables fine-grained, policy-driven authentication across the delivery lifecycle. In later sections, we preview how these identities can be used for access control, justification-aware authorization, and intent-based governance.

As enterprises increasingly adopt Zero Trust principles, the management of Non-Human Identities (NHIs)—such as CI/CD runners, service accounts, and automation agents—has become a top priority. Gartner recently highlighted NHI management as a 2025 strategic trend, underscoring the need for policy-aware identity issuance and governance for ephemeral workloads \cite{gartner-nhi}.

\section{Historical Approaches to CI/CD Authentication}

\subsection{Static Secrets}

Legacy authentication in CI/CD systems was centered around long-lived credentials, often stored as encrypted pipeline variables or pulled from external secrets managers like HashiCorp Vault. These credentials included cloud IAM tokens, API keys, and service account passwords, all of which were injected into jobs at runtime. While operationally simple, this model suffers from multiple shortcomings:

\begin{itemize}
  \item \textbf{Over-permissioning:} Credentials are frequently scoped for broad access, violating least privilege.
  \item \textbf{Lack of traceability:} Pipelines reuse identities across environments, branches, and repositories, making attribution difficult.
  \item \textbf{Inflexibility:} Secret rotation or revocation requires manual intervention across pipelines and teams.
  \item \textbf{No contextual awareness:} Secrets cannot express why access is needed or under what justification.
\end{itemize}

Even with dynamic secrets from tools like Vault, the authentication model remains static. Workloads and jobs are not attested at runtime, and the identity tied to a credential does not change with the pipeline’s context. In multi-tenant environments, shared runners and static roles can lead to credential leakage or unintended cross-tenant access.

\textbf{Example: Injected static credentials}

\begin{verbatim}
export AWS_ACCESS_KEY_ID="AKIA***********"
export AWS_SECRET_ACCESS_KEY="****************"
\end{verbatim}

These secrets often persist in logs or shell history and are difficult to audit. The limitations of this model led to the rise of identity federation approaches like OpenID Connect (OIDC).

\subsection{OIDC Federation}

OpenID Connect (OIDC) has become a widely adopted approach for CI/CD-to-cloud authentication. Instead of injecting secrets, CI platforms like GitHub Actions, GitLab CI, and Azure DevOps issue ephemeral tokens containing signed claims about the job context. Cloud providers validate these tokens and issue short-lived credentials using trust relationships like AWS STS, GCP Workload Identity Federation, or Azure federated credentials.

\textbf{Example: GitHub Actions to AWS STS}

\begin{verbatim}
{
  "Effect": "Allow",
  "Principal": {
    "Federated": "arn:aws:iam::ACCOUNT_ID:oidc-provider/token.actions.githubusercontent.com"
  },
  "Action": "sts:AssumeRoleWithWebIdentity",
  "Condition": {
    "StringLike": {
      "token.actions.githubusercontent.com:sub": "repo:org/repo-name:ref:refs/heads/main"
    }
  }
}
\end{verbatim}

Each repo or workflow must be explicitly mapped to a role. As environments scale, policy drift and management overhead increase.

\textbf{Example: Decoded GitHub-issued OIDC token (excerpt)}

\begin{verbatim}
{
  "sub": "repo:org/my-service:ref:refs/heads/main",
  "aud": "sts.amazonaws.com",
  "job_workflow_ref": "org/my-service/.github/workflows/deploy.yml@refs/heads/main",
  "repository": "org/my-service",
  "sha": "6cffe1e4b9ea91dd0a189..."
}
\end{verbatim}

GitHub uses \texttt{job\_workflow\_ref} for unique job identity. GitLab, in contrast, provides \texttt{CI\_JOB\_ID}, a numeric value. Azure federated credentials are scoped to GitHub environment bindings. These variations reflect the lack of standardization across platforms.

Despite their improvements, OIDC federation models present challenges:

\begin{itemize}
  \item \textbf{Policy sprawl:} Role mappings for each job or repo become hard to scale and audit.
  \item \textbf{Claim fragility:} Tokens are tied to claims like \texttt{sub} and \texttt{ref}, which break with pipeline restructuring.
  \item \textbf{Tooling gaps:} Cloud providers lack visibility into federated identity usage or access posture.
  \item \textbf{Lack of abstraction:} No unified standard exists across cloud identity federation models.
\end{itemize}

OIDC federation improves CI/CD identity posture within platform-bound contexts but falls short in multi-cloud and federated environments. It does not offer a portable, runtime-verifiable identity primitive that decouples trust from the platform itself. This is the problem SPIFFE aims to solve.

\section{Limitations of Current Patterns}

Although OIDC federation and secret management solutions have improved CI/CD security, they still fall short of delivering verifiable, runtime-scoped identity that aligns with Zero Trust principles. The limitations of current authentication models become more visible in multi-cloud and enterprise environments.

\begin{itemize}
    \item \textbf{Static Role Bindings:} Federated trust relationships rely on pre-defined mappings between workflows and cloud IAM roles. These bindings are brittle, hard to maintain, and difficult to scale as teams and pipelines grow.
    
    \item \textbf{Lack of Intent Awareness:} OIDC tokens and injected credentials cannot express \textit{why} access is requested. This makes it difficult to enforce justification-based policies or perform meaningful audit correlation.

    \item \textbf{No Approval Integration:} Traditional pipelines lack hooks for human or automated approval gating that ties into authentication. This breaks the link between access requests and organizational policy enforcement.

    \item \textbf{Poor Federation Across Clouds:} Each cloud provider implements OIDC federation differently, with unique claims and trust models. This fragmentation makes it hard to establish consistent access patterns across hybrid or multi-cloud architectures.

    \item \textbf{Unclear Separation of Identity and Authorization:} In many systems, identity and access rights are bundled together. This tight coupling limits flexibility and prevents enforcement of fine-grained authorization policies.
\end{itemize}

These gaps create an opportunity for a stronger identity foundation—one that can issue workload identities at runtime, travel with the job or workload, and support policy-driven access control across environments. SPIFFE addresses this need directly.

These shortcomings also reflect gaps noted in NIST SP 800-204, which calls for runtime identity and authorization in microservices-based systems \cite{nist-sp800204}. SPIFFE extends these recommendations to the CI/CD layer, bringing verifiable, workload-scoped identity to modern pipelines.

\section{SPIFFE as a Standard for Workload Identity}

The Secure Production Identity Framework for Everyone (SPIFFE) defines a set of open standards for issuing, representing, and verifying identities for workloads. Unlike traditional identity systems that rely on static credentials or platform-bound tokens, SPIFFE enables the issuance of cryptographically verifiable identity documents at runtime, based on the workload’s environment and attestation data.

\begin{figure}[h]
  \centering
  \rotatebox{270}{\includegraphics[width=0.7\textwidth]{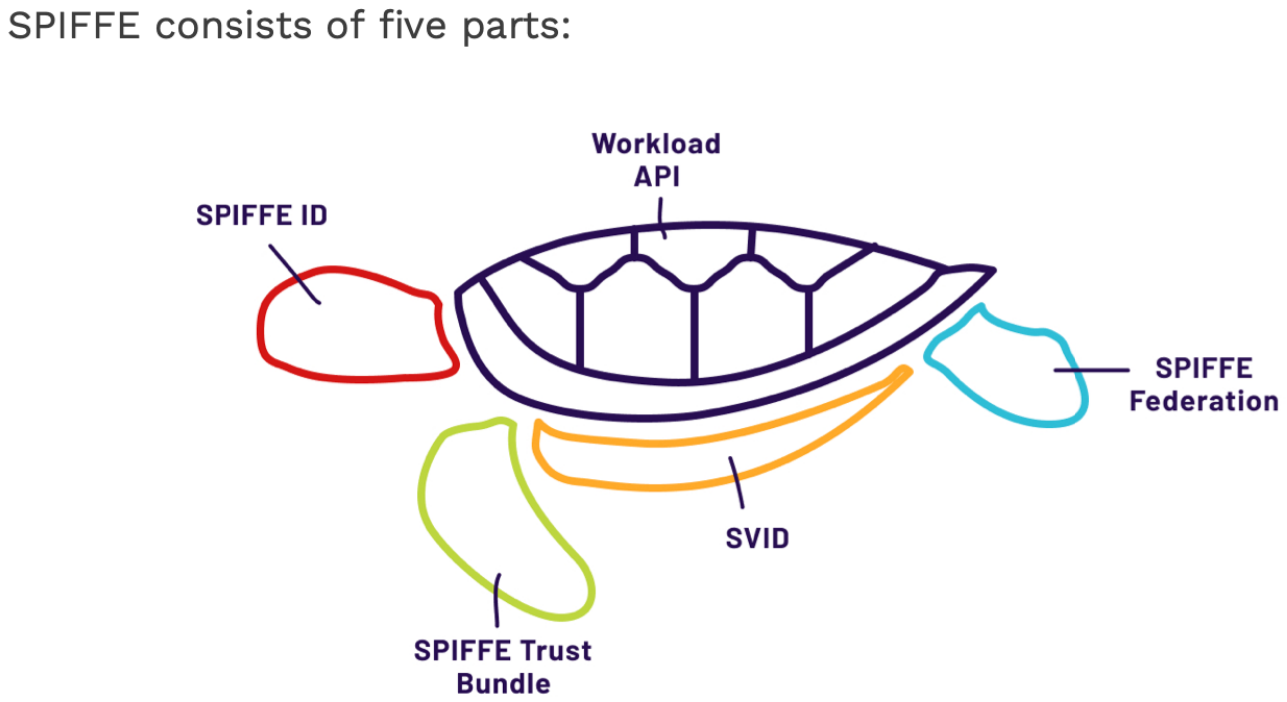}}
  \caption{The five core components of the SPIFFE identity model. Source: \textit{Solving the Bottom Turtle}, licensed under CC BY 4.0.}
  \label{fig:spiffe-five-parts}
\end{figure}

Figure~\ref{fig:spiffe-five-parts} provides a visual overview of the SPIFFE identity model. While this paper does not cover each component in depth, the diagram serves as a useful orientation for readers interested in exploring the full specification and implementation landscape. SPIFFE defines five fundamental elements that work together to establish secure workload identity:

\begin{itemize}
    \item \textbf{SPIFFE ID:} A URI-formatted identifier that uniquely names a workload or service within a trust domain.
    \item \textbf{SVID (SPIFFE Verifiable Identity Document):} A cryptographically verifiable document—X.509 or JWT—that proves possession of a SPIFFE ID.
    \item \textbf{Workload API:} A node-local interface through which workloads retrieve their identities securely at runtime, without needing to authenticate first.
    \item \textbf{Trust Bundle:} A collection of root or intermediate CA public keys used to validate SVIDs issued by a SPIFFE authority.
    \item \textbf{Federation:} A trust model that allows multiple SPIFFE trust domains to interoperate by exchanging trust bundles, enabling cross-domain workload authentication.
\end{itemize}

\subsection{What Is a SPIFFE ID?}

A SPIFFE ID is a unique, URI-based identity assigned to a workload. It follows the format:

\begin{verbatim}
spiffe://<trust-domain>/<path>
\end{verbatim}

For example: \texttt{spiffe://org.example/frontend/build-runner}

The trust domain defines the administrative and cryptographic boundary for SPIFFE identity issuance. Each workload receives a SPIFFE ID that reflects its logical role, and this identity is stable across restarts, deployments, or network changes.

\begin{figure}[h]
  \centering
  \includegraphics[width=0.7\textwidth]{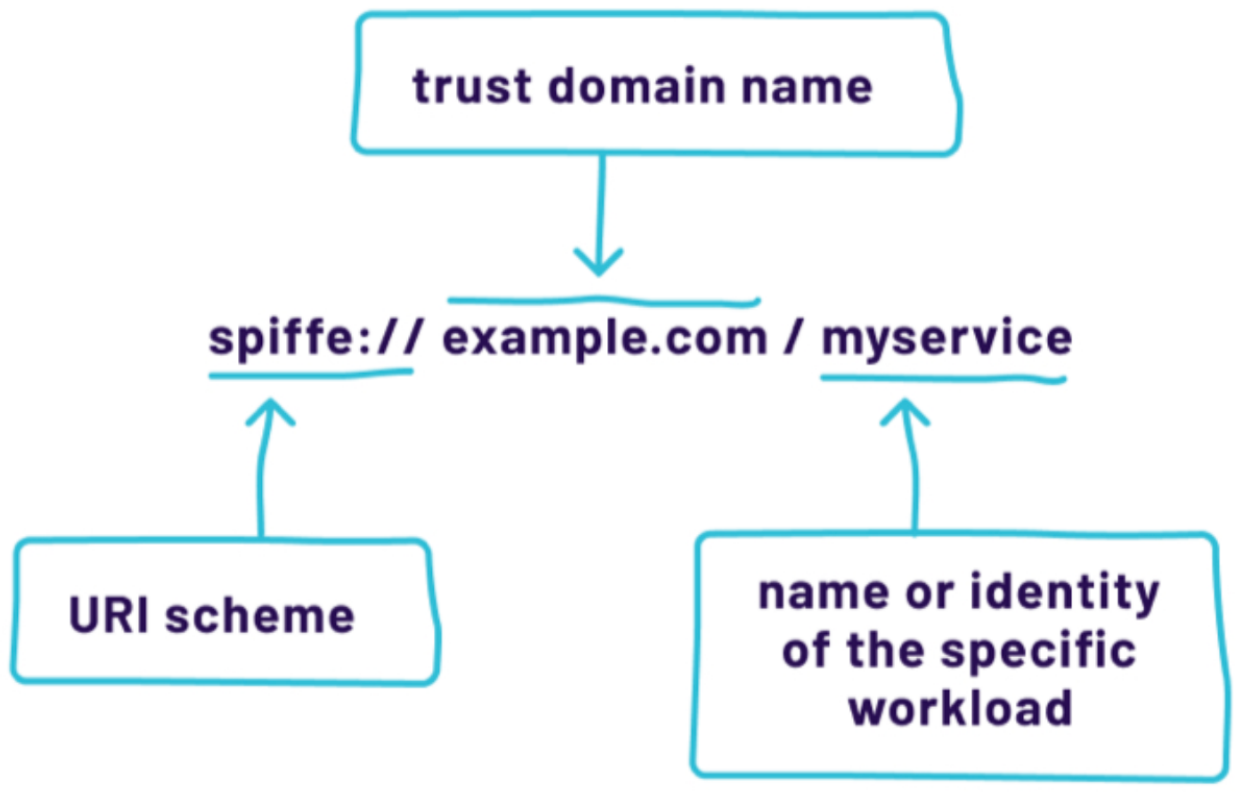}
  \caption{Structure of a SPIFFE ID and its mapping to workloads within a trust domain. Source: \textit{Solving the Bottom Turtle}, licensed under CC BY 4.0.}
  \label{fig:spiffe-id-structure}
\end{figure}

Figure~\ref{fig:spiffe-id-structure} illustrates the structure of a SPIFFE ID and how it maps to workloads within a trust domain.

\subsection{Workload Attestation}

In SPIFFE, identity is not granted statically or manually. Instead, every workload must undergo \textit{attestation}—a process by which the SPIRE Agent verifies environmental attributes before issuing a SPIFFE ID. This ensures that identities are tied to real, verifiable conditions at runtime.

SPIRE performs two layers of attestation:
\begin{itemize}
    \item \textbf{Node attestation:} Verifies the identity of the host running the SPIRE Agent, using metadata from the cloud instance, TPMs, or other platform-bound mechanisms.
    \item \textbf{Workload attestation:} Validates attributes such as Kubernetes service accounts, container image hashes, or filesystem paths.
\end{itemize}

Attestation is handled locally by the SPIRE Agent using \textbf{selectors}, which extract metadata from the host platform. These selectors can include:

\begin{itemize}
    \item Kubernetes service account, namespace, and pod labels
    \item Executable binary path
    \item Docker container runtime metadata
    \item Host-specific data such as EC2 tags or node UUIDs
\end{itemize}

The SPIRE Server uses these selectors to match the workload against registration policies. If the selectors match a valid entry, the Agent issues an SVID through the Workload API.

\begin{figure}[h]
  \centering
  \rotatebox{270}{\includegraphics[width=0.7\textwidth]{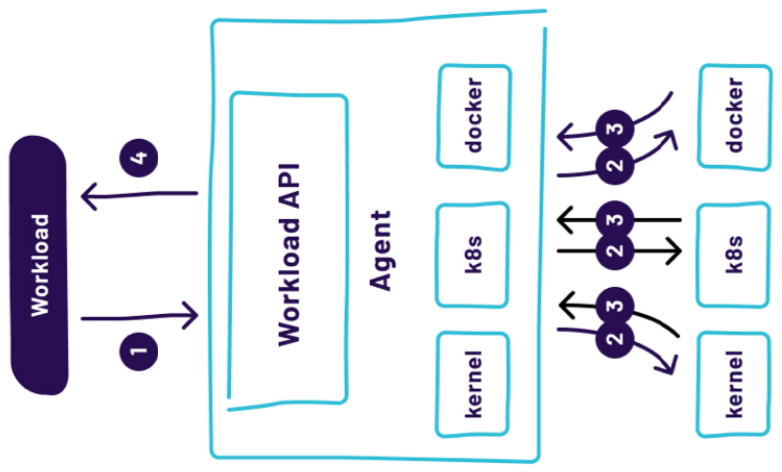}}
  \caption{Workload attestation flow in SPIRE using platform-specific plugins and the Workload API. Source: \textit{Solving the Bottom Turtle}, licensed under CC BY 4.0.}
  \label{fig:workload-attestation}
\end{figure}

Figure~\ref{fig:workload-attestation} shows how the Agent performs attestation using different plugins—such as kernel, Kubernetes, or Docker—to derive selectors. Once verified, it returns a valid identity through the Workload API.

This dynamic, plug-in-based attestation is critical for ephemeral environments like CI/CD pipelines, where jobs must be verified and issued identity at runtime, without hardcoded trust or pre-registered secrets.

\subsection{Verifiable Workload Identity: SVIDs and SPIRE}

SPIFFE Verifiable Identity Documents (SVIDs) are the cryptographic credentials used to prove workload identity. These are issued dynamically at runtime and come in two main formats:

\begin{itemize}
    \item \textbf{X.509-SVIDs:} Used for mutual TLS (mTLS) authentication in workload-to-workload communication. The SPIFFE ID is embedded as a URI in the certificate’s Subject Alternative Name (SAN) field.
    \item \textbf{JWT-SVIDs:} Used for OIDC-compliant identity federation in cloud environments, HTTP-based APIs, or service meshes that consume bearer tokens.
\end{itemize}

For example, a JWT-SVID used in a cloud authentication flow may look like:

\begin{verbatim}
{
  "aud": "sts.amazonaws.com",
  "sub": "spiffe://ci/org/deploy-job",
  "exp": 1717200000,
  "iat": 1717196400,
  "iss": "spiffe://org.example"
}
\end{verbatim}

This token conforms to OIDC expectations and can be passed to identity-aware services like AWS STS using the \texttt{AssumeRoleWithWebIdentity} API to obtain scoped credentials at runtime.

To use a SPIFFE-issued JWT with AWS, an IAM role must be configured to trust the SPIRE trust domain as an identity provider. A sample trust policy might look like:

\begin{verbatim}
{
  "Version": "2012-10-17",
  "Statement": [
    {
      "Effect": "Allow",
      "Principal": {
        "Federated": "arn:aws:iam::ACCOUNT_ID:oidc-provider/spire.example.org"
      },
      "Action": "sts:AssumeRoleWithWebIdentity",
      "Condition": {
        "StringEquals": {
          "spire.example.org:sub": "spiffe://ci/org/deploy-job"
        }
      }
    }
  ]
}
\end{verbatim}

This OIDC-compliant policy ensures that only jobs with the correct SPIFFE ID—issued by SPIRE at runtime—can assume the role. It enables least privilege access without relying on long-lived credentials or CI platform-specific tokens.

SPIRE (SPIFFE Runtime Environment) automates this lifecycle—handling attestation, SVID issuance, rotation, and trust bundle distribution. It consists of a central SPIRE Server and one or more node-local SPIRE Agents that interact with workloads.

\subsection{Trust Domains and Federation}

SPIFFE introduces the concept of trust domains to isolate administrative boundaries and cryptographic trust. Cross-domain federation is enabled through trust bundles — collections of public keys that allow workloads in one domain to authenticate workloads in another, without requiring global trust anchors.

Trust bundles define the root CA material used to validate SVIDs across SPIFFE deployments. Federation enables trust between independently managed domains—critical in CI/CD pipelines that span vendors, clouds, or business units.

This trust model makes it possible for CI/CD jobs, build runners, and deployed workloads to participate in identity federation securely — even across organizational or cloud boundaries.

SPIFFE federation follows IETF draft-ietf-spiffe-federation v07 \cite{ietf-spiffe}, using X.509 bundle exchange over HTTPS with mTLS authentication.

\subsection{Comparison with OIDC Federation}

Unlike OIDC federation, SPIFFE operates entirely within the workload runtime and is not tied to a CI/CD provider or cloud platform. Identity is issued based on attested attributes like the workload’s binary path, Kubernetes service account, or node metadata. This enables strong identity guarantees, even in ephemeral environments like CI jobs or containers.

OIDC tokens are scoped to platform-defined job metadata and require pre-configured IAM roles and trust policies. SPIFFE IDs, in contrast, are portable, runtime-verifiable, and usable across heterogeneous infrastructure. They enable identity-centric policy enforcement and eliminate reliance on static or platform-bound credentials.

\section{Applying SPIFFE in CI/CD Workflows}

CI/CD systems often operate with broad privileges but lack strong, verifiable identity. Secrets and static roles are still common, and OIDC federation is tied to the source platform. SPIFFE offers a runtime-issued identity model that improves isolation, traceability, and policy enforcement across the pipeline.

\subsection{Identity for CI Job Runners}

Ephemeral CI jobs—such as GitHub runners or Kubernetes jobs—can be attested at runtime based on selectors like image hashes, service accounts, or host attributes. Once attested, the SPIRE Agent issues a SPIFFE ID that reflects the job’s purpose (e.g., \texttt{spiffe://ci/org/release-job}).

In multi-tenant CI/CD environments, this pattern is even more powerful. For example, a runner in a centralized CI platform shared across teams might receive an identity such as:

\begin{verbatim}
spiffe://platform.example.org/ci/team-a/release-runner
\end{verbatim}

This ID clearly reflects the platform boundary (\texttt{platform.example.org}), tenant (\texttt{team-a}), and workload role (\texttt{release-runner}). It enables scoped access control, policy enforcement, and audit tracking without relying on hardcoded roles or credentials.

This pattern works across a variety of environments. SPIRE can issue identities to Kubernetes jobs, GitHub-hosted runners using self-managed agents, or even bare-metal CI agents running on hardened Linux hosts. Identity issuance is driven by attestation plugins, not tied to platform-specific capabilities.

Each SPIRE Agent issues SVIDs based on local attestation, but does not require direct permissions to assume cloud roles. Instead, a job-specific JWT-SVID can be passed to AWS STS (or a similar cloud identity service), which validates the SPIFFE ID via federation and issues short-lived credentials. This enables per-job authorization across accounts or environments—without granting persistent privileges to shared agents or relying on static, cross-tenant IAM roles.

This pattern is especially valuable in multi-cloud or multi-account environments where CI/CD platforms serve multiple teams. Each job receives an isolated identity, mapped to least-privilege permissions via policy, with no need for runners to be bound to privileged service accounts or shared cloud roles.

\subsection{Identity for Deployment Targets}

Deployed workloads also receive SPIFFE IDs based on their environment. This enables mutual authentication: the CI job has its identity, and the runtime workload has its own. The two can authenticate each other using mTLS or token verification, allowing secure, traceable handoffs.

\subsection{Cross-Domain Federation}

In multi-team or multi-cloud environments, SPIFFE’s trust domain model supports identity federation across boundaries. Workloads and CI jobs from different domains (e.g., \texttt{team-a.example.org}, \texttt{platform.example.org}) can trust each other through exchanged trust bundles—without breaking isolation or over-permissioning.

This architecture mitigates several common CI/CD risks: compromised runners can no longer impersonate privileged identities; cross-tenant privilege escalation is blocked by scoped trust domains; and credentials are not exposed in logs or artifacts. SPIFFE-based identity reduces the impact of supply chain attacks by enforcing identity at runtime, rather than relying on static job metadata or pre-provisioned tokens.

These identities serve as the basis for just-in-time, policy-driven access decisions, which we explore in the next section.

\section{Benefits of SPIFFE-Based Workload Identity in CI/CD}

Replacing secrets with SPIFFE-based identity brings several tangible benefits to CI/CD systems—especially those operating in multi-team, multi-cloud, or regulated environments.

First, the system becomes \textbf{secrets-free}. There is no need to inject or manage long-lived credentials in job configurations, runner environments, or external secrets managers. This reduces the blast radius of credential leaks and removes the operational burden of secret rotation and lifecycle tracking.

Second, SPIFFE supports \textbf{runtime-issued, verifiable identity}. Workloads and jobs are identified based on who they are, not where they run or how they were triggered. This shifts trust from environment-based assumptions to runtime attestation, allowing identity to be tightly bound to selectors like service account, image hash, and host metadata.

Third, SPIFFE IDs are \textbf{platform-agnostic and policy-compatible}. They can be consumed by policy engines such as Open Policy Agent (OPA), Cedar, or custom ABAC frameworks to enforce fine-grained access controls. Identity becomes decoupled from the CI provider or infrastructure role mapping, allowing policy logic to focus on intent and function.

Fourth, SPIFFE enables organizations to \textbf{enforce least privilege} in CI/CD. Each job can be independently scoped to access a minimal set of resources—cloud APIs, registries, secrets—based on its unique SPIFFE ID. Policies can be driven by workload role rather than by shared execution context or static job labels.

Fifth, SPIFFE facilitates \textbf{auditability and access justification}. Each identity is traceable, renewable, and scoped to a runtime context. This supports compliance programs and enables integration with systems that enforce human approvals or SLA-driven access gates.

Finally, SPIFFE IDs form the \textbf{foundation for workload-based authorization}. SPIFFE provides identity but does not directly confer access—this role is delegated to a policy-based system that maps identities to short-lived credentials or capabilities. This separation of identity and access is what enables dynamic, Zero Trust enforcement, which we explore in the following section.

These properties lay the groundwork for scalable, just-in-time, and intent-aware identity enforcement in CI/CD workflows.

Table~\ref{tab:comparison} summarizes how SPIFFE compares to traditional CI/CD identity models across key dimensions of trust, scope, and portability.

\begin{table}[h]
\centering
\caption{Comparison of authentication models for CI/CD workload identity}
\label{tab:comparison}
\renewcommand{\arraystretch}{1.3}
\begin{tabular}{|l|c|c|c|}
\hline
\textbf{Feature}               & \textbf{Static Secrets} & \textbf{OIDC Federation} & \textbf{SPIFFE (w/ SPIRE)} \\
\hline
Credential Injection           & Yes                     & No                        & No                         \\
Runtime Issuance               & No                      & Partial                   & Yes                        \\
Platform Neutral               & No                      & No                        & Yes                        \\
Identity Portability           & Low                     & Medium                    & High                       \\
Supports Federation            & No                      & Limited                   & Yes                        \\
Tied to Job Context            & No                      & Yes                       & Yes                        \\
Supports mTLS Authentication  & No                      & No                        & Yes                        \\
\hline
\end{tabular}
\end{table}

\vspace{1em}
In addition to improving portability and policy enforcement, SPIFFE also mitigates key security risks commonly encountered in CI/CD environments:

\begin{itemize}
  \item \textbf{Credential Leaks:} Short-lived SVIDs eliminate the need for static secrets in runners or job environments.
  \item \textbf{Privilege Escalation:} Trust domain boundaries restrict identity issuance and prevent cross-tenant impersonation.
  \item \textbf{Replay Attacks:} JWT-SVIDs include expiration, issuer, and audience fields to support bounded, verifiable access.
  \item \textbf{Zero Trust Alignment:} SPIFFE enforces runtime authentication, continuous verification, and least-privilege access by design.
\end{itemize}

\section{From Identity to Access: Enabling Policy-Based Authorization}

SPIFFE provides strong identity—but identity alone does not grant access. CI/CD systems must translate that identity into scoped permissions such as access to cloud APIs, internal secrets, or deployment targets. This transition is governed by policy.

A policy-based access system serves as the decision point between a verified identity and a protected resource. When a CI job or workload presents its SPIFFE ID, a policy engine—such as Open Policy Agent (OPA) or Cedar—evaluates whether access should be granted, to which resource, under what conditions, and for how long. This replaces static role bindings with real-time, context-aware decision making.

\subsection*{Example: OPA Rego Policy}

\begin{verbatim}
package authz

default allow = false

allow {
  input.spiffe_id == "spiffe://ci/org/deploy"
  input.action == "write"
  input.resource == "s3://prod-release-artifacts"
}
\end{verbatim}

This policy allows write access to a release bucket only for jobs bearing a specific SPIFFE ID. The evaluation is scoped to the identity, resource, and action.

\subsection*{Example: Cedar Policy}

Cedar expresses access control as a set of entities, actions, and policies:

\begin{verbatim}
permit(
  principal == workload::"spiffe://ci/org/deploy",
  action == action::"publish",
  resource == artifact::"release-bucket"
);
\end{verbatim}

This policy grants the `publish` action to the deploy job’s identity on the release artifact resource. Unlike role-based systems, both policies enforce access based on verifiable identity, not environmental assumptions.

This architecture enables just-in-time access, audit logging, and runtime enforcement. Access can be gated by human approvals, bound to job metadata, or tied to SLA-compliant runtime conditions.

\subsection*{Formalization}

Formally, identity-based access can be represented as:

\[
\text{AccessGranted} = f(\text{SPIFFE\_ID}, \text{Context}, \text{Policy})
\]

Where \textit{Context} may include workload metadata, time constraints, tags, justification tokens, or prior approvals.

This model supports dynamic access control in CI/CD pipelines where identity is issued at runtime and access decisions are driven by policy. In the next paper, we explore how this model can be extended to include credential issuance, intent validation, and continuous policy supervision.

\section{Related Work and Standardization Landscape}

SPIFFE and SPIRE are part of a broader effort to standardize non-human identity across distributed systems. This movement intersects with cloud-native access control, Zero Trust architectures, and software supply chain integrity initiatives.

The \textbf{SPIFFE/SPIRE project}, a graduated CNCF project, defines the identity primitives used in this paper. SPIRE supports Kubernetes-native attestation, multi-platform workload selectors, and pluggable federation with third-party PKIs. Its architecture makes it a practical implementation of runtime-issued identity.

\textbf{GitHub Actions} and other CI platforms support OIDC-based federation with AWS, GCP, and Azure. These mechanisms reduce reliance on static secrets, but are tightly bound to platform-specific claims and workflow metadata. They lack portability across providers and often do not support multi-domain federation.

The IETF \textbf{WIMSE (Workload Identity Management in Secure Environments)} working group is drafting specifications for SPIFFE federation and workload identity portability.\footnote{\url{https://datatracker.ietf.org/wg/wimse/}} These drafts formalize trust bundle exchange and claims-based authorization logic between independent domains.

Other complementary standards include:
\begin{itemize}
    \item \textbf{SCITT (Secure Component Identification and Transparency)}: An IETF initiative to ensure software component trust throughout the supply chain.\footnote{\url{https://datatracker.ietf.org/doc/draft-ietf-scitt-architecture/}}
    \item \textbf{SLSA (Supply Chain Levels for Software Artifacts)}: A framework for build and release integrity, often used alongside SPIFFE to bind provenance to identity.\footnote{\url{https://slsa.dev}}
\end{itemize}

Legacy models—such as HashiCorp Vault, Kubernetes secrets, and static IAM roles—remain common but assume implicit trust and are ill-suited for federated, multi-tenant environments. These models lack runtime verification and do not scale to ephemeral CI/CD workloads or cross-cloud governance.

SPIFFE distinguishes itself by acting as a vendor-neutral, runtime identity layer that can integrate with OIDC, credential brokers, or SBOM frameworks. These developments point toward a converging architecture: portable, verifiable identity as the foundation for cloud-native authorization and Zero Trust delivery pipelines.

Recent industry adoption further highlights the shift toward verifiable, brokered workload access. In April 2024, Snowflake publicly described its implementation of a Workload Identity and Access Management (Workload IAM) model using Aembit.\footnote{\url{https://medium.com/snowflake-builders/securing-workload-access-snowflakes-journey-to-workload-iam-1c2b3b5f51f6}} Their approach mirrors the principles explored in this paper: runtime identity, dynamic credential issuance, and policy-based access across cloud and SaaS environments. SPIFFE serves as a foundational layer in such systems, decoupling identity issuance from cloud platforms and enabling secretless authentication via policy brokers.

\section{Conclusion and Future Directions}

In modern CI/CD workflows, identity is the foundation for access, compliance, and trust. Yet most systems still rely on secrets, environment-based assumptions, or tightly coupled OIDC tokens issued by the platform itself. These models fail to scale across cloud accounts, tenants, or ephemeral workloads.

This paper introduced SPIFFE as a portable, runtime-issued identity model that enables CI jobs and deployed workloads to authenticate securely without relying on static credentials. By decoupling identity from infrastructure, SPIFFE makes it possible to establish mutual authentication, enforce least privilege, and drive access decisions through policy.

We showed how SPIFFE can be applied to ephemeral CI runners, deployment targets, and multi-tenant environments. We explored how workload identity forms the foundation for secretless, policy-driven access via brokers or authorization engines like OPA and Cedar. These patterns not only improve security, but also reduce operational burden, enable automation, and support audit and compliance.

As CI/CD pipelines grow more dynamic and cross-organizational, SPIFFE serves as a unifying control-plane primitive. It replaces brittle secrets with cryptographic identities, and shifts access enforcement to policy systems that evaluate who, why, and under what conditions access is allowed.

In future work, we explore how these identities can be used to enforce access through broker systems, dynamic credential issuance, intent-aware governance, and justification-driven workflows. The second paper in this series focuses on credential brokers, and the third on runtime authorization and compliance control loops.

\appendix

\section*{Appendix}

\subsection*{A. Example SPIFFE IDs}

\begin{verbatim}
spiffe://org.example/frontend/build-runner
spiffe://platform.example.org/ci/team-a/release-runner
\end{verbatim}

These IDs reflect trust domain boundaries and job roles, enabling scoped access and federated identity.

\subsection*{B. Sample SPIRE Registration Entry}

\begin{verbatim}
entry {
  spiffe_id = "spiffe://org.example/frontend/build-runner"
  selector = "k8s_sa:build"
  parent_id = "spiffe://org.example/spire/agent/k8s-node"
  ttl = 3600
}
\end{verbatim}

This entry binds a Kubernetes service account to a SPIFFE ID using workload attestation.

\subsection*{C. Comparison Matrix}

\begin{center}
\renewcommand{\arraystretch}{1.3}
\begin{tabular}{|l|c|c|c|}
\hline
\textbf{Feature}               & \textbf{Static Secrets} & \textbf{OIDC Federation} & \textbf{SPIFFE (w/ SPIRE)} \\
\hline
Credential Injection           & Yes                     & No                        & No                         \\
Runtime Issuance               & No                      & Partial                   & Yes                        \\
Platform Neutral               & No                      & No                        & Yes                        \\
Identity Portability           & Low                     & Medium                    & High                       \\
Supports Federation            & No                      & Limited                   & Yes                        \\
Tied to Job Context            & No                      & Yes                       & Yes                        \\
Supports mTLS Authentication  & No                      & No                        & Yes                        \\
\hline
\end{tabular}
\end{center}

\subsection*{D. Reference Architecture Snippet}

\begin{itemize}
    \item \textbf{SPIRE Server:} High-availability deployment, backed by etcd or PostgreSQL
    \item \textbf{SPIRE Agents:} Deployed as Kubernetes DaemonSet or on CI runners
    \item \textbf{Selectors:} k8s\_sa, docker\_label, aws\_iam\_tag
    \item \textbf{Trust Federation:} Configured with trust bundle exchange
\end{itemize}

\subsection*{E. References and Tooling Links}

\begin{itemize}
    \item \textbf{SPIFFE Specification:} \url{https://github.com/spiffe/spiffe}
    \item \textbf{SPIRE Runtime:} \url{https://github.com/spiffe/spire}
    \item \textbf{SPIFFE IETF Draft:} \url{https://datatracker.ietf.org/doc/draft-ietf-spiffe-federation/}
    \item \textbf{WIMSE WG @ IETF:} \url{https://datatracker.ietf.org/wg/wimse/}
    \item \textbf{SLSA Framework:} \url{https://slsa.dev}
    \item \textbf{Aembit Workload IAM:} \url{https://aembit.io}
    \item \textbf{Snowflake Builders Blog:} \url{https://medium.com/snowflake-builders/securing-workload-access-snowflakes-journey-to-workload-iam-1c2b3b5f51f6}
\end{itemize}

\end{document}